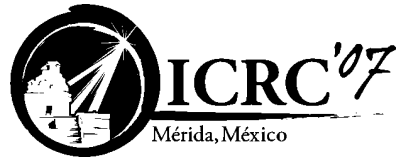
ICRC'07
Mérida, México

# Calibration of the CREAM-I calorimeter

Y.S. YOON[1,2], H.S. AHN[2], M.G. BAGLIESI[3], G. BIGONGIARI[3], O. GANEL[2], J.H. HAN[2], J.A. JEON[4],
K.C. KIM[2], M.H. LEE[2], L. LUTZ[2], P. MAESTRO[3], A. MALININ[2], P.S. MARROCCHESI[3], S. NAM[4],
I.H. PARK[4], N.H. PARK[4], E.S. SEO[1,2], R. SINA[2], J. WU[2], J. YANG[4], R. ZEI[3], S.Y. ZINN[2]
[1]*Dept. of Physics, University of Maryland, College Park, MD 20742 USA*
[2]*Inst. for Phys. Sci. and Tech., University of Maryland, College Park, MD 20742 USA*
[3]*Dept. of Physics, University of Siena and INFN, Via Roma 56, 53100 Siena, Italy*
[4]*Dept. of Physics, Ewha Womans University, Seoul, 120-750, Republic of Korea*
ysy@physics.umd.edu

**Abstract:** The Cosmic Ray Energetics And Mass (CREAM) calorimeter is designed to measure the spectra of cosmic-ray particles over the energy range from ~$10^{11}$ eV to ~$10^{15}$ eV. Its first flight as part of the CREAM-I balloon-borne payload in Antarctica during the 2004/05 season resulted in a record-breaking 42 days of exposure. Calorimeter calibration using various beam test data will be discussed in an attempt to assess the uncertainties of the energy measurements.

## Introduction

The CREAM calorimeter was designed to measure the energy of cosmic-ray nuclei in the range ~$10^{11}$–$10^{15}$ eV [1,2]. To correctly measure energy over this wide range, calibration is quite important. After the initial calibration [3,4], further corrections were implemented. Using the updated calibration constants, various test beam data were compared with Monte Carlo (MC) simulation data. The systematic calibration uncertainty was assessed based on this comparison, as well as on beam test data with different energies in different time.

## The Calorimeter Calibration Process

The CREAM calorimeter calibration is based on identifying for each event the ribbon in each layer having the highest signal in that layer. By comparing the maximum ribbon signal in each layer with MC simulations, a calibration constant in MeV/ADC units is obtained (see [3,4] for more details). Although the beam spot position is known with respect to the ribbons, one wishes to exclude those events where the incident particle hit a neighboring ribbon, so only those events are selected where the ribbon nominally "in the beam" actually records the highest signal. In general this correctly selects the appropriate events. Several improvements have recently been implemented to this calibration process, including coherent noise correction and hit selection with normalized gain.

## Coherent Noise Correction

In some events, noise pickup can affect all channels of one application specific integrated circuit (ASIC) in a similar manner, giving rise to coherent behavior. By studying the behavior of channels with no optical signal input, such coherent behavior can be identified, and its effect greatly reduced. This is done by measuring the change in the "monitor channel" relative to the mean value of its pedestal distribution, and correcting, on an event-by-event basis, the electronic pedestal values for channels reading out optical signals for that event by the same amount. Before applying this correction, the ADC sum plots for several calorimeter channels showed broad or distorted pedestal and signal distributions. Following application of the above correction, the pedestal and signal peaks of these distributions became narrower and more Gaussian in shape. The majority of channels, where such coherent behavior was not observed, showed little impact from the procedure, with only a small (and expected) increase in pedestal width, due to the increase in the number of channels introduced by the correction.

**Event Selection using Normalized Gain**

Although the ribbon hit by the beam particle normally has the highest signal, different channels have different light yield, light collection and light transmission efficiencies, are read out with different hybrid photo diodes (HPDs) having different quantum efficiencies and gains, with different ASIC gains, etc. Thus, it is possible that the ribbon with highest signal would have a lower overall gain, to the point of not recording the largest signal in the layer for most events. To correct for this selection bias and increase the sample of events used in the calibration, one needs to modify the selection, by making a less stringent requirement than that the ribbon signal be higher than all other ribbon signals in the layer for that event. This modified selection was implemented as follows. For each ribbon k, the mean signal, $\mu_k$, was calculated using the original selection

(1)     $S_k > S_n$  for  n = 1 – 50; n ≠ k

where $S_n$ is the signal in the nth ribbon. Selection factors were then calculated for the resulting distributions.

(2)     $\alpha_k = <\mu_k> / \mu_k$

where $\mu_n$ is the mean of the signal distribution for the nth ribbon. A new selection was then applied.

(3)     $\alpha_k \times S_k > \alpha_n \times S_n$  for  n = 1 – 50; n ≠ k

This modified selection assures that where a channel has an especially low gain relative to its neighbors, $\alpha_k$ would be larger than $\alpha_n$, resulting in more of the events where the energy deposit in the ribbon "in the beam" are highest being selected. Once this selection is imposed, selecting more of the correct events, the mean value is calculated for the uncorrected signals, thus avoiding any bias in the gain calibration constants. This process resulted in significant improvement for several ribbons, with minimal impact on the remaining the rest of ribbons.

## Confirming the Calibration

The calibration constants were calculated for the CREAM-I calorimeter by comparing data from X and Y scans with 150 GeV/c electron beams to MC simulations based on GEANT/FLUKA 3.21 [5,6]. This calibration was verified by applying the calibrations derived from this procedure to other data-sets such as electron beams of other energies, proton beams, and heavy ion beams.

**Electron Energy Scan**

Calibrated energy sums were plotted for electron beam events with energies of 50, 100, 150 and 200 GeV/c, incident on the central region of the calorimeter, and compared to MC distributions with electronic noise and photon statistics implemented in the simulation. It shows good linearity for different energies. See more detail results [3,4]

**350 GeV/c Protons**

During the beam test, proton beam data with 250 GeV/c and 350 GeV/c were collected. These events were simulated with the same conditions as the beam runs. To compare only those events similar to those of interest in flight, only well-contained events with significant shower activity were selected. Figure 1 shows fairly good agreement between beam data and simulations using these cuts.

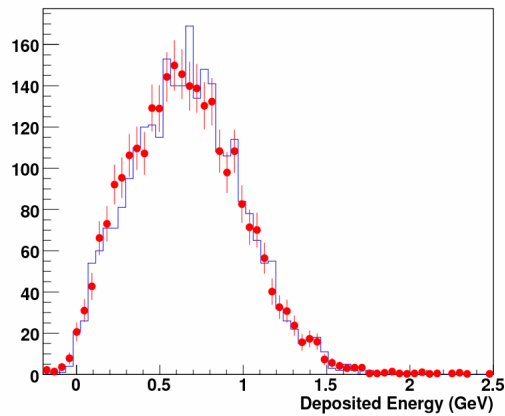

Figure 1: Deposited energy in 350 GeV/c proton beam (red circles) and MC events (histogram). Electronics noise and photon statistics were implemented in the MC. A cut on sum of deposited energy in several layers was applied to both beam and MC samples to select interacting particles.

**Heavy Ion Beam data**

During the beam test, data were collected from an A/Z = 2 nuclear fragment beam from a 158 GeV/A Indium beam incident on a thin target. Since the energy of the fragments is linear in A (and thus in Z, for A/Z = 2), the response of the calorimeter to particles with higher charge and higher energy could be tested. Using the Silicon Charge Detector (SCD) [7], each particle was identified and its deposited energy measured by the calorimeter. At each Z value, events in ± 1σ of the Gaussian fit to the charge were selected, and their mean deposited energy was obtained by Gaussian fit. Figure 2 shows the correlation between mean energy and mass. The plot shows good linearity up to A = 58, implying that the calorimeter is linear at least up to 9.2 TeV (58 × 158 GeV/c = 9.16 TeV/c), showing the offset is less than 1% at 8.8 TeV.

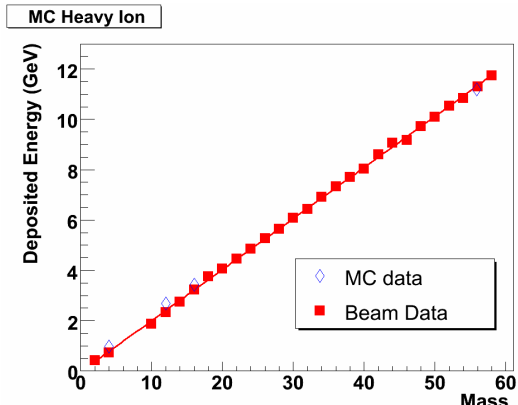

Figure 2. Energy deposit from A/Z = 2 nuclear fragments produced by a 158 GeV/A indium beam incident on a thin target. Beam data (red triangles) after calibration are consistent with MC (blue circles) generated for Z = 2, 6, 8 and 56.

Figure 3 is a comparison of the deposited energy between the beam data identified as oxygen nuclei and MC data of vertically incident 2528 GeV oxygen using FRITIOF/RQMD [8, 9] interfaced to the GEANT/FLUKA 3.21 hadronic simulation package. The figure shows good agreement around the peak. According to the calculation using material densities, 90% of oxygen should interact in the carbon targets and the calorimeter. Figure 3 shows a small peak near zero, representing ~10% non-interacting events, and it is consistent with the calculation. In the beam data, heavier nuclei (Z>8 in this case) may interact upstream, and be identified as oxygen. These could appear as background, leading to excess in the tails of the distribution.

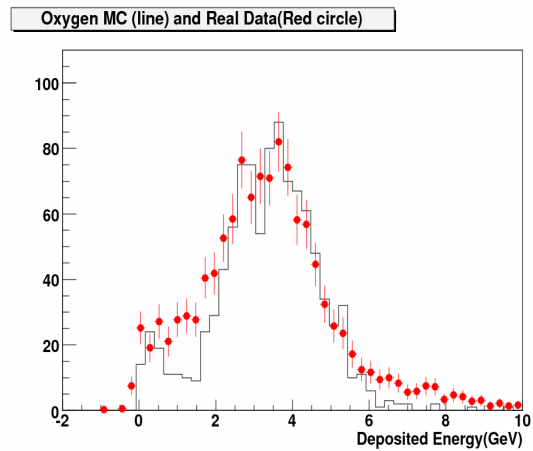

Figure 3. Deposited energy from events identified as oxygen using an SCD cut (7.5 < Z < 8.5) (red circles) and MC simulation results (histogram).

**Estimating Systematic Uncertainty**

Several components contribute to the systematic uncertainty in the CREAM calorimeter energy measurement.

The calibration process described above accounts for most major corrections needed. This includes gain correction due to different light yields, light collection and transport efficiencies, high voltage values of different supplies at the time of calibration, HPD quantum efficiencies, ASIC gains, dead channel corrections, etc.

Other factors still remain that could potentially affect the energy reconstruction. These include, e.g., HPD gain changes due to different HV values relative to the calibration run, any replacement of HPDs that change the quantum efficiency, temperature dependence of the readout electronics, the exact level of extrapolation accuracy to energies far above the beam energies available, etc. Calibration was carried out with 150 GeV/c electron beams. These were compared with 50 GeV/c beam scan data taken in about 2 months after the calibration run. Figure 4 shows a narrow distribution about the expected ratio of 3.0, show-ing linearity at low energy. As seen above, this linearity extends at least up to 9.2 TeV. Above this energy the linearity can only be estimated using MC with lab test results of electronics uniformity and the uncertainty in the process of "stitching" the low-, mid- and high-energy readout ranges.[10]

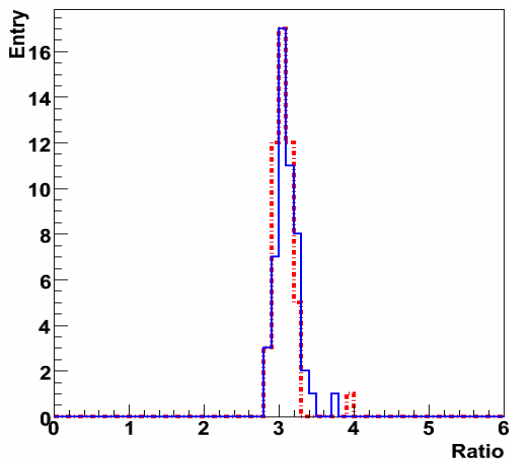

Figure 5. Ratio of signal sums between 50 and 150 GeV/c electrons before (blue solid line) and after (red dashed lihe) calibration for 50 different beam injection points. Uniformity of response improved from 4.8% to 3.8% after calibration.

## Conclusions

After improving calibration constants by applying coherent noise and normalized gain corrections, various data sets were tested. 350 GeV/c proton data show excellent agreement with simulation results, confirming that calibration with 150 GeV/c electrons works very well for proton measurements. The heavy ion data shows good linearity up to 9.2 TeV after calibration. Comparing 150 GeV/c and 50 GeV/c electron scan data shows <2% offset from the expected ratio of 3.0 and <4% uncertainty.

## Acknowledgements

This work is supported by NASA, INFN, KICOS, and MOST and CSBF. The authors also thank CERN for the excellent beam test facilities and operations.